\documentclass[twocolumn,english,aps,prl,showpacs,nofootinbib]{revtex4}
\usepackage[T1]{fontenc}
\usepackage[latin9]{inputenc}
\usepackage{amsmath}
\usepackage{amssymb}
\usepackage{esint}
\usepackage{graphicx}
\usepackage{verbatim}
\usepackage{chngcntr}
\usepackage[section]{placeins}
\usepackage{lipsum}
\makeatletter
\usepackage{float}

\@ifundefined{textcolor}{}
{%
 \definecolor{BLACK}{gray}{0}
 \definecolor{WHITE}{gray}{1}
 \definecolor{RED}{rgb}{1,0,0}
 \definecolor{GREEN}{rgb}{0,1,0}
 \definecolor{BLUE}{rgb}{0,0,1}
 \definecolor{CYAN}{cmyk}{1,0,0,0}
 \definecolor{MAGENTA}{cmyk}{0,1,0,0}
 \definecolor{YELLOW}{cmyk}{0,0,1,0}
 }

\usepackage{nicefrac}\usepackage{dsfont}
\usepackage[normalem]{ulem}\@ifundefined{definecolor}
 {\usepackage{color}}{}

\long\def\/*#1*/{}

\makeatother

\usepackage{babel}
\begin{document}

\title{Anomalous discontinuity at the percolation critical point of active gels}

\author{M. Sheinman\textsuperscript{1,2}, A. Sharma\textsuperscript{1}, J. Alvarado\textsuperscript{3,4}, G. H. Koenderink\textsuperscript{3}, F. C. MacKintosh\textsuperscript{1}}
\address{\textsuperscript{}Department of Physics and Astronomy, VU University, Amsterdam, The Netherlands\\
\textsuperscript{2}Max Planck Institute for Molecular Genetics, 14195 Berlin, Germany\\
\textsuperscript{3}FOM Institute AMOLF, Science Park 104, 1098 XG Amsterdam, The Netherlands\\
\textsuperscript{4}Department of Mechanical Engineering, Hatsopoulos Microfluids Laboratory, Massachusetts Institute of Technology, Cambridge, Massachusetts 02139, United States}

\date{\today}
\begin{abstract}
We develop a percolation model motivated by recent experimental studies of gels with active network remodeling by molecular motors.
This remodeling was found to lead to a critical state reminiscent of random percolation (RP), but with a cluster distribution inconsistent with RP. 
Our model not only can account for these experiments, but also exhibits an unusual type of mixed phase transition: 
We find that the transition is characterized by signatures of criticality, but with a discontinuity in the order parameter.
\end{abstract}
\maketitle

Percolation theory has become pervasive in a number of fields ranging from Physics to Mathematics and even Computer Science \cite{sahimi1994applications}. In particular, it successfully describes connectivity and elastic properties of polymer networks \cite{de1979scaling,broedersz2014modeling}. 
The simplest percolation model is the random percolation (RP) model, consisting of a collection of nodes with controlled connectivity, $ p $, representing the fraction of occupied bonds between the nodes. As a function of $p$, the order parameter---the mass fraction of the largest cluster---becomes finite above the percolation threshold $p_c$. The nature of the transition is of special interest because the system properties are highly tunable at this point, especially if the transition is discontinuous; in that case, just a few bonds can have a significant impact, even for very large systems \cite{nagler2011impact}. 
Usually, however, percolation transitions are second-order, with a continuous variation of the order parameter and various critical signatures. More specialized percolation models can exhibit different phase behavior, including discontinuous transitions between the two phases (see discussion below).

Here, we present a simple model based on random percolation that develops a discontinuous jump in the order parameter in the thermodynamic limit, while exhibiting other features of criticality in such quantities as the correlation length and susceptibility. Interestingly, the transition we observe occurs for the same $p_c<1$ as for random percolation. Moreover, our model can account for recent experimental results on active biopolymer gels that have been shown to self-organize towards a critical connectivity point \cite{alvarado2013myosin}. The experimentally observed cluster properties at this point were found to be \emph{inconsistent} with the ordinary random percolation model. 

\begin{figure}[t]
\centering
  \begin{tabular}{@{}cccc@{}}
  \includegraphics[width=.25\textwidth]{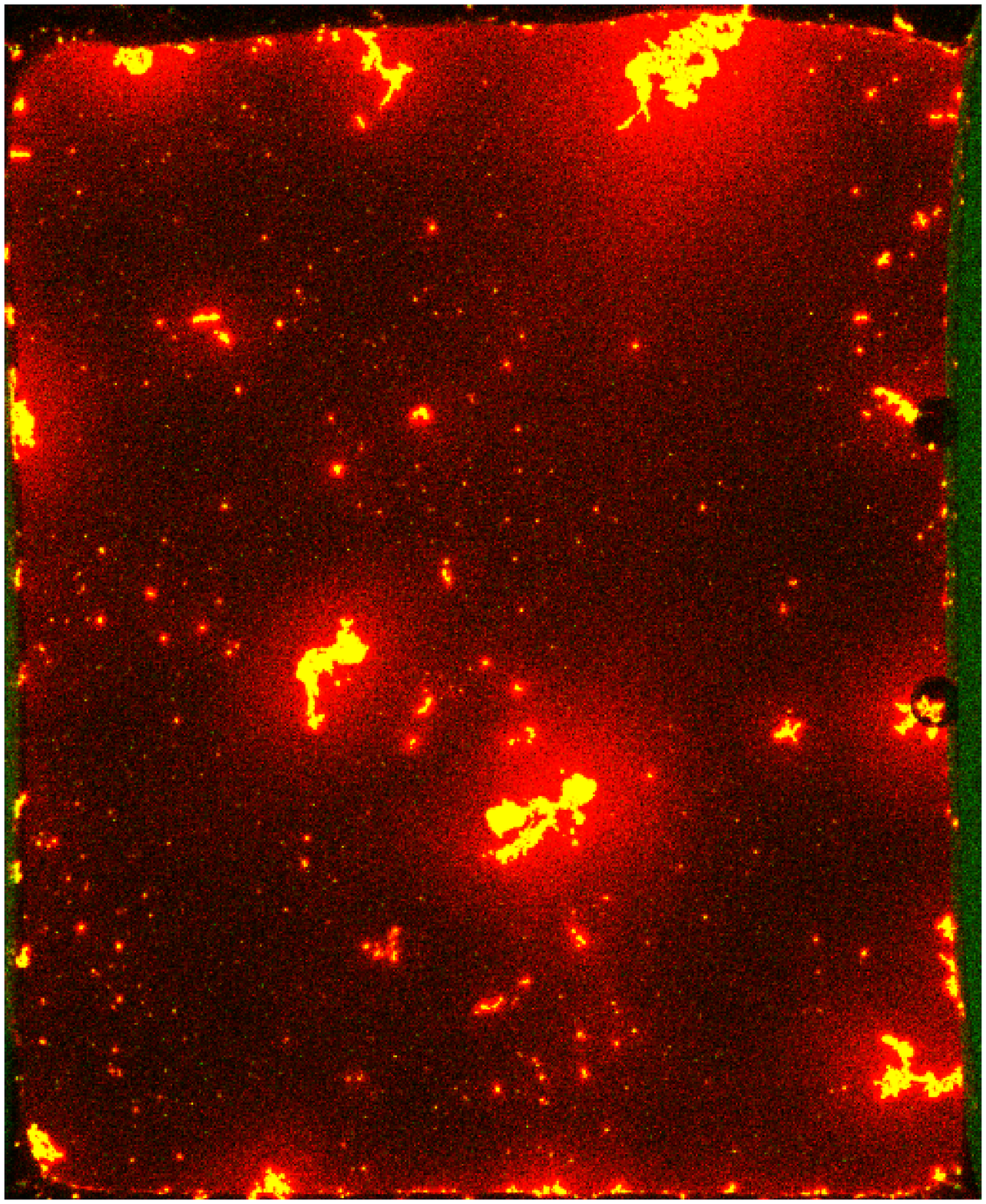}
    \put(-130,147){\fcolorbox{white}{white}{(a)}} 
    &
    \includegraphics[width=.25\textwidth]{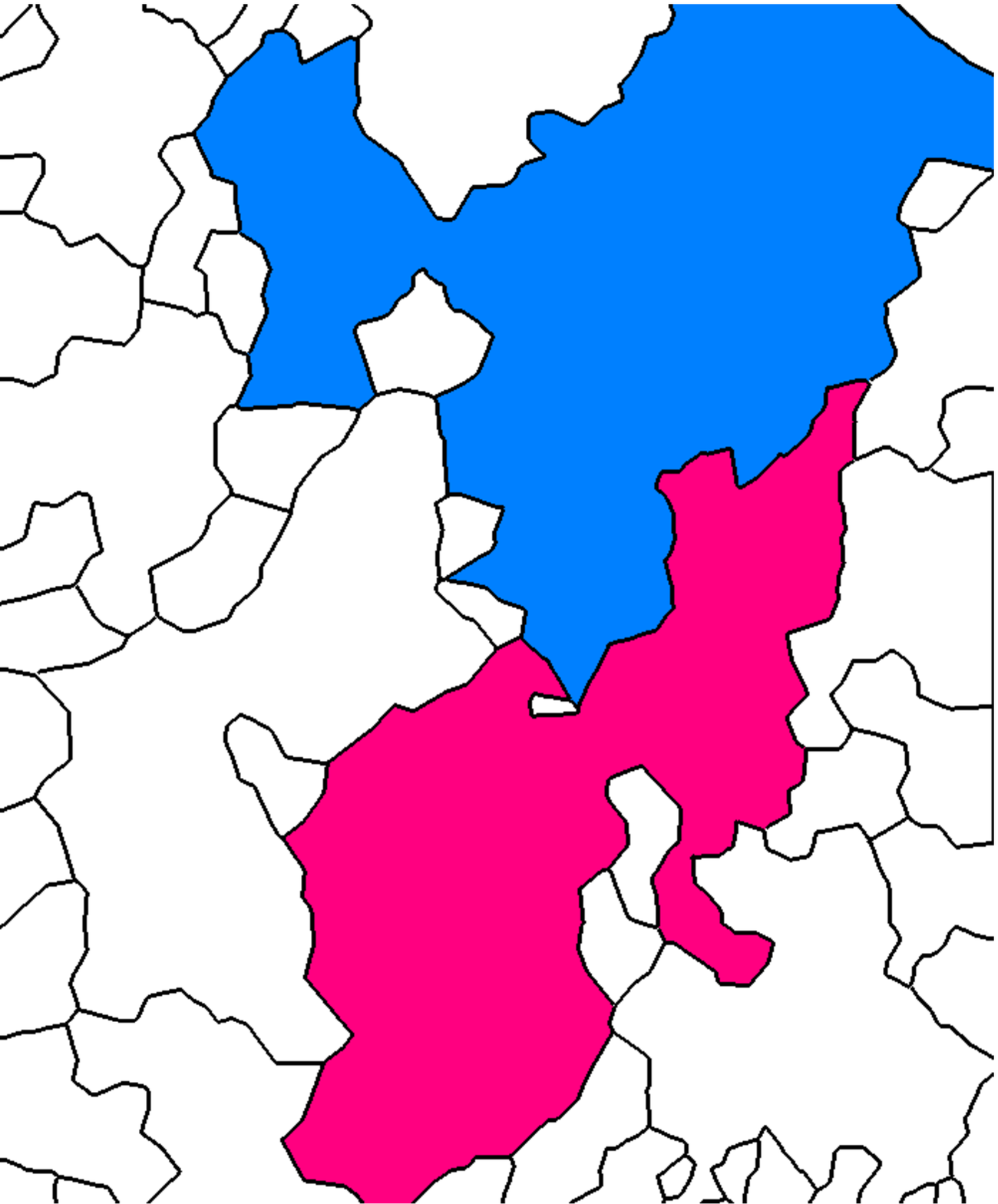}
    \put(-132,147){\fcolorbox{white}{white}{(b)}} 
    \\
    \includegraphics[width=.25\textwidth]{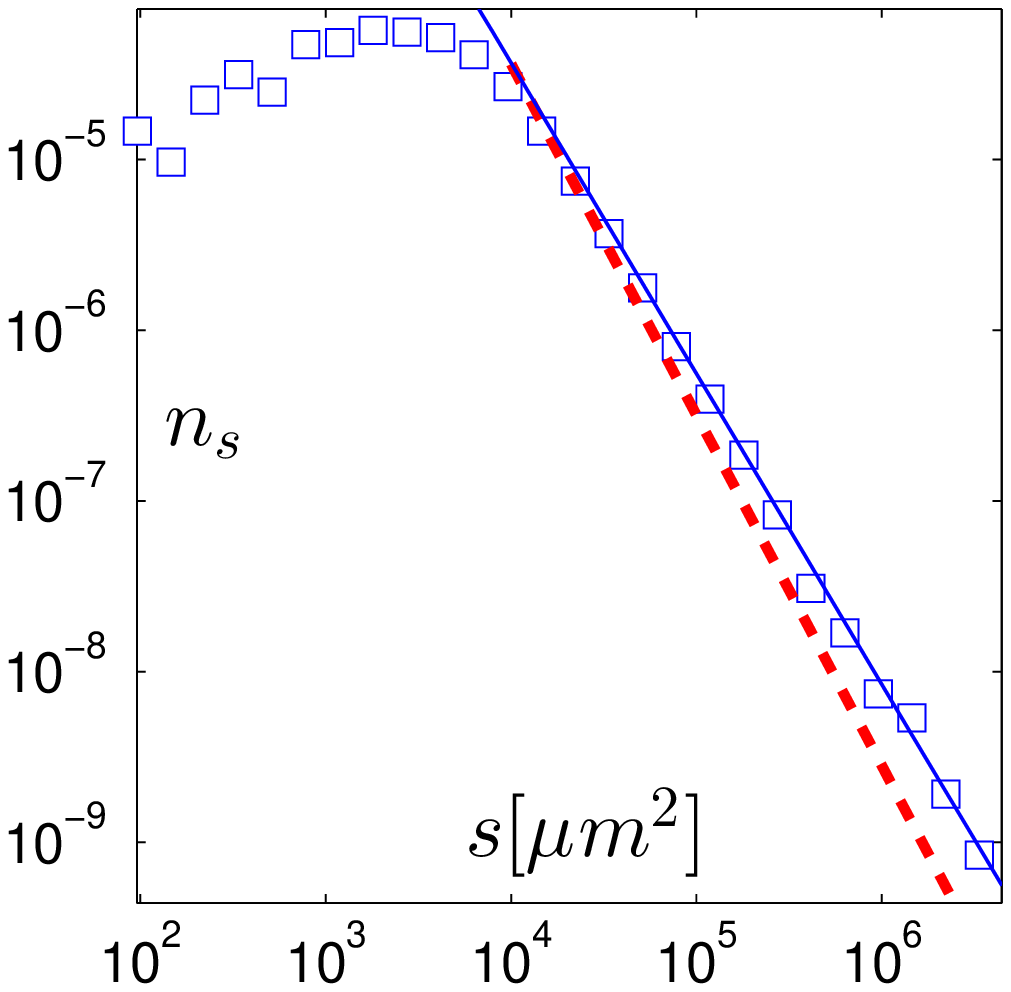}
    \put(-125,120){{(c)}}
    \put(-40,83){\includegraphics[width=.07\textwidth]{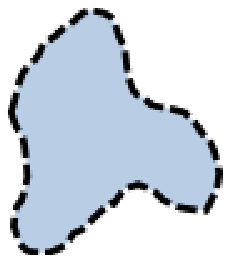}} 
    \put(-90,40){\includegraphics[width=.07\textwidth]{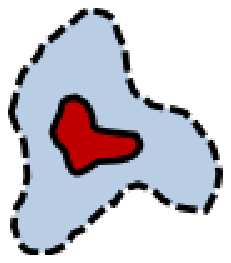}} 
    &
    \includegraphics[width=.25\textwidth]{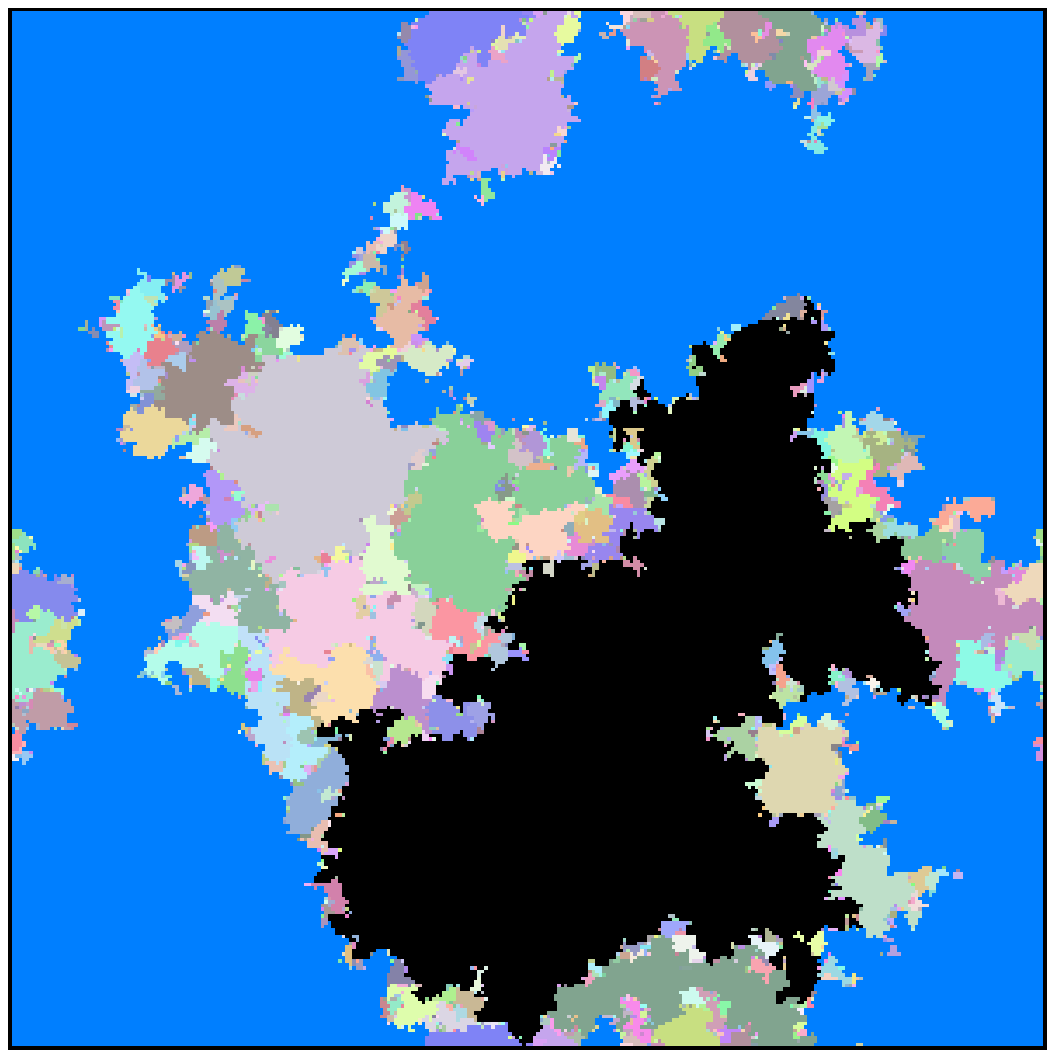}
    \put(-125,118){(d)}
  \end{tabular}
\caption{(Color online) (a) Experimental results of a fascin-crosslinked actin network, collapsed by myosin motors  (see SI for details and movie of the collapse). (b) Initial configuration of the collapsed clusters in (a). Colours indicate the largest (blue) and the second-largest (pink) clusters. (c) Histogram (squares) of cluster masses, averaged over 26 samples. For the critically connected regime, the data is statistically more consistent ($1.4$ standard errors from the Hill estimator of $\tau=1.91 \pm 0.06$ \cite{clauset2009power}) with a power-law distribution with a NEP model's Fisher exponent from Eq.~\eqref{tau} (solid line). The agreement of the data with the RP model Fisher exponent $\tau'=187/91$, indicated by the dashed line, is significantly worse ($2.4$ standard errors from the Hill estimator). Lower insets demonstrate an enclave (solid line) and its surrounding cluster (dashed line). In the upper inset the enclave is absorbed into its surrounding cluster. (d) Clusters' structure  at the critical point of the NEP model (RP model with absorbed enclaves). The absorption of enclaves is implemented by identifying the longest boundary of each cluster and absorbing all the nodes within this boundary in the cluster (see SI for more details).}
\label{Jose}
\end{figure}

In these experiments, we studied a model cytoskeletal system, composed of actin filaments, fascin cross-links and myosin motors in a quasi-2D chambers of dimensions 3mm$\times2$mm$\times80\mu$m \cite{alvarado2013myosin} (see SI). We observed a motor-driven collapse of the network into disjointed clusters (see Fig.\ \ref{Jose}(a) and movie of the collapse in SI). The configuration of the clusters prior the collapse is obtained by analyzing the time-reversed movie (see Fig.\ \ref{Jose}(b)) and their masses were estimated from their initial areas. We found that over a wide range of the experimental parameters, the number $n_s$ of clusters of mass $s$ exhibit a power-law distribution: $n_s \sim s^{-\tau}$. Here, $\tau$ is the Fisher power-law exponent, which must be strictly limited to values $\geq2$ for the RP model \cite{stauffer1994introduction}. It was found experimentally, however, that $\tau\simeq 1.91\pm0.06$. A key feature of these experiments is the apparent absence of \emph{enclaves}---clusters fully surrounded by another cluster (see inset of Fig.\ \ref{Jose}(c)). These enclaves are responsible for the fractal nature of clusters and $\tau>2$ in the RP model \cite{den1979relation,nienhuis1982exact}. Below we show that the, apparently paradoxical, experimental features of $n_s \sim s^{-\tau}$, yet with $\tau<2$ can be understood within our no-enclaves percolation (NEP) model.

\begin{figure}
\centering
    \includegraphics[width=\columnwidth]{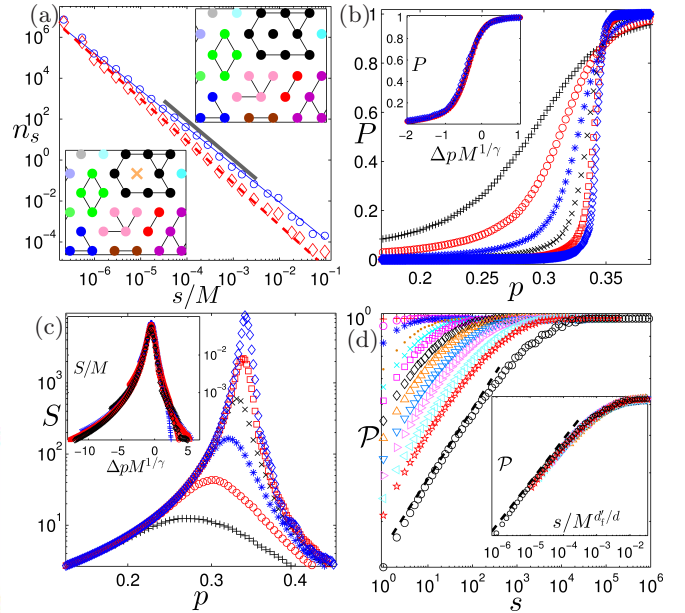} 
        \put(-240,223){(a)}
        \put(-123,223){(b)}
        \put(-238,105){(c)}
        \put(-120,100){(d)}
\caption{(Color online) (a) Cluster mass distribution at the percolation transition of the RP (diamonds) and NEP (circles) models, with $ M=2000^2$. The long lines indicate power-laws with $ \tau'=187/91 $ (dashed line) and $ \tau=1.82 $ (solid line) (Eq.~\eqref{tau}). The short line indicates the experimental $\tau \simeq 1.91$ observed over 2 decades. Insets illustrate cluster structures of the RP  (left inset) and NEP models (right inset) on a triangular lattice for $0<p<1$ (see SI for $p=0,1$ cases). Colors (online) represent different clusters. The cross in the left inset indicates an enclavic cluster that is absorbed in the NEP model (right inset). The absorption of enclaves in the NEP model changes the distribution of clusters sizes and can account for the experimental observation of $\tau<2$ in Ref.\ \cite{alvarado2013myosin}. (b) Demonstration of the discontinuous transition in the NEP model. Strength of the largest cluster, $P$, as a function of bond density, $p$ for different system sizes, from $M=12$ to $400$. Inset: collapse of the data from the main figure using Eq.\ \eqref{Pcollapse} and $\gamma=8/3$. (c) Demonstration of the critical transition in the NEP model. Average cluster, $S$, as a function of bonds density, $p$ for the same system sizes as in (b). Inset: collapse of the data from the main figure using Eq.\ \eqref{Scollapse} and $\gamma=8/3$. (d) Probability of an $ s $-cluster to be \emph{not} an enclave of another cluster, $\mathcal{P}$, at the percolation transition of the RP model for different system sizes, from $M=6$ to $2000$. Inset: collapse of the data from the main figure using Eq.~\eqref{P}. The dashed lines indicate power-law $ a=0.1706$ obtained from Eqs.~\eqref{a} and \eqref{tau}.}
\label{NEP4}
\end{figure}

\textit{NEP model and its theoretical analysis}---Our NEP model begins with the \emph{random} percolation model, but in which enclaves---clusters fully surrounded by a larger cluster---are absorbed into the surrounding cluster (see Figs.\ \ref{Jose}(c),\ref{NEP4}(a) and SI). We do this because in the experiments, during the collapse, as a single connected cluster contracts, it tends to incorporate other material within the cluster, including distinct enclaves contained within it (see illustration in SI). Thus, experimentally the final cluster configuration, obtained from the time-reversed movie, is enclave-free (see Fig.\ \ref{Jose}(b)).
In the NEP model, as in the RP model, massive nodes are located on a regular lattice. Nearest-neighbor nodes are connected by massless bonds with a probability $p$. For any $p$, the total system mass (in units of nodes) is equal to the total number of nodes, $M$.
For a given value of $ p $, the network is obtained from the corresponding network of the RP model, but with all enclaves absorbed into their surrounding clusters (they can also be absorbed during each step of the dilution protocol). This is illustrated in the insets to Figs. \ref{Jose}(c) and \ref{NEP4}(a) and in SI.
As we show, our NEP model exhibits mixed properties of both discontinuous and continuous phase transitions, including the anomalous critical behavior, consistent with the experiments. 

In the 2D \emph{random} percolation model, close to the percolation transition point, $p=p_c$, the strength (i.e., fraction of total mass) of the cluster with the largest mass scales as $P \sim \left| \Delta p \right|^{\beta'}$, where $\Delta p=p-p_c$. $P$ vanishes at the critical point in the thermodynamic limit, such that the largest cluster mass scales as $M^{d_\mathrm{f}'/d}$. In the following, we denote by primed symbols the quantities for the RP model and reserve unprimed symbols for the NEP model. The transition in the RP model is critical, such that the correlation length diverges as
$
\xi \sim \left| \Delta p \right|^{-\nu'},
$
 with $\nu'=4/3 $ \cite{den1979relation,nienhuis1982exact}.
The cluster masses are distributed as a power-law with a Fisher exponent $\tau'=187/91>2$:
$
n_s' \sim M s^{-\tau'},
$
where $n_s'$ is the number of clusters with mass $s$ (see Fig.\ \ref{NEP4}(a)). The Fisher exponent is related to the fractal dimension by the hyperscaling relation, $\tau'=d/d_\mathrm{f}'+1$. Therefore, the continuity of the transition in an RP model follows from $d_\mathrm{f}'<d$ or, equivalently, from $\tau'>2$. Thus, the qualitative form of the percolation transition is reflected in the value of the Fisher exponent and fractal dimension \cite{stauffer1994introduction}.

In contrast, the absorption of enclaves in the NEP model leads to a different universality class, although the critical value $p_c$ remains the same, since enclaves do not themselves percolate. Moreover, because enclaves are wholly surrounded by another cluster, the absorption of enclaves does not change the scaling of the radius of gyration of clusters. Thus, the divergence of the correlation length is as in the RP model, with $\nu=\nu'$. After absorption of enclaves, the surviving compact clusters are euclidean with $d_\mathrm{f}=d=2$.

The transition in the NEP model is still critical, since the correlation length diverges there. Therefore, the distribution of cluster masses is scale-free at the transition. The mass of the largest cluster below the transition point, $p<p_c$,  scales as $\xi^d$, such that it remains finite in the thermodynamic limit, $M \rightarrow \infty$. At the transition point, where $\xi \sim M^{1/d}$, the mass of the largest cluster scales as $M$. Thus, the strength of the largest cluster in the thermodynamic limit exhibits a discontinuous jump from $0$ for $p<p_c$ to a positive value at $p=p_c$. 
The only possibility to get both the discontinuity and the criticality at the transition point is that the number of clusters with mass $s$ is sublinear with the system size and is given by
\begin{equation}
n_s \sim M^{\tau-1} s^{-\tau},
\label{ns}
\end{equation}
with Fisher exponent $\tau<2$. Only in this case the critical, power-law distribution possesses a sufficiently heavy tail such that the mass of the largest cluster scales as the total mass, providing both criticality and discontinuity (see Refs.~\cite{Friedman2009Construction,da2010explosive,hooyberghs2011criterion,da2014Critical} and SI). Other exceptions to the usual second-order nature of percolation transition include interdependent networks~\cite{buldyrev2010catastrophic,gao2011networks}, hierarchical structures \cite{boettcher2012ordinary}, which can exhibit discontinuous transitions between the two phases.
 The possibility of such first-order-like percolation transitions in simple networks has been the subject of considerable recent debate. Starting from the explosive percolation \cite{achlioptas2009explosive}, this and other models have been analyzed that show sharp transitions~\cite{moreira2010hamiltonian,cho2010cluster,d2010local,radicchi2010explosive,da2010explosive,araujo2010explosive,riordan2011explosive,nagler2012continuous,da2014Critical,da2014Solution}, which, in some cases, nevertheless become continuous in the thermodynamic limit. Mixed phase transitions were found for several update percolation procedures~\cite{araujo2011tricritical}. In some cases the transition is at the point $p_c=1$ (see, e.g., Ref.~\cite{cho2013avoiding}) while in other cases it is for $p_c<1$ (like in the NEP model and, e.g., Ref.~\cite{schroder2013crackling}).
 Also bootstrap percolation models have been found to exhibit discontinuous phase transitions, although it remains unclear whether these are critical in 2D \cite{chalupa1979bootstrap,schwarz2006onset,Chae2014Complete}. 
Beyond percolation, similar issues arise, e.g., in thermal systems \cite{thouless1969long,kafri2000why,bar2014mixed} and in the jamming transition, for which several phase-defining quantities or order parameters are possible. Interestingly, some of these exhibit discontinuous behavior, while others are continuous \cite{van2010jamming}.

The upper bound of $2$ for the Fisher exponent in the NEP model can be obtained analytically. To do so, consider the scaling of number of clusters possessing mass $s$ at the transition point, $n_s \left( p=p_c \right)$. A cluster in the NEP model with no enclaves is euclidean in contrast to the RP model where clusters possess fractal dimension $d_\mathrm{f}'$. Therefore, one can relate $n_s$ and the same quantity of the RP model, $n_s'$, using
\begin{equation}
n_s \sim n_{s'}'  \mathcal{P}\left(s'\right)  \frac{d s'}{ds} \sim 
M s^{-2} 
\mathcal{P}\left(s'\right) .
\label{nsP}
\end{equation}
Here $s'$ is $s^{d_\mathrm{f}'/d}$ and $\mathcal{P}\left(s\right)$ is the probability that a cluster with mass $s$ is \textit{not} an enclave in the RP model. The last equality in Eq.\ \eqref{nsP} follows from the hyperscaling relation of $\tau'$ and $d_\mathrm{f}'$.
Smaller clusters are expected to have a higher probability to be an enclave and, therefore, to be absorbed in their surrounding clusters during the elimination of enclaves. Thus, $\mathcal{P}\left(s\right)$  monotonically increases with $s$. This implies an upper bound for the Fisher exponent of the NEP model, $\tau<2$, in agreement with the experiment.
For such a heavy tail distribution of cluster masses the total number of clusters, $n$, is not an extensive quantity. Integrating Eq.\ \eqref{ns}, one obtains 
\begin{equation}
n \sim M^{\tau-1}.
\label{n}
\end{equation}

In the NEP model the size of the jump of $P$ at the transition and the scaling ansatz for $ P $ can now be obtained using Eq.\ \eqref{ns} with $ \tau<2 $. For $p>p_c$ the largest cluster has a mass that scales as $M$. The remaining mass of all the other clusters combined scales sublinearly with $M$, as $\int^{\xi^2} n_s s ds \sim M^{\tau-1} \xi^{4-2\tau}$. Thus, the strength of the largest cluster is $1$ above the transition in the thermodynamic limit. In this limit the discontinuity of $P$ in the NEP model is from $0$ to $1$ at $p=p_c$. The finite-size scaling ansatz for the largest cluster strength is given by (see SI)
\begin{equation}
P=\mathcal{G}\left( \Delta p M^{1/\gamma}\right),
\label{Pcollapse}
\end{equation}
where $\mathcal{G}(-\infty)=0$ and $\mathcal{G}(\infty)=\Delta P=1$.

One can calculate other critical exponents of the NEP model in the following way.
Since all the clusters are euclidean, $ d_\mathrm{f}=d=2 $, the cutoff of the $n_s$ power-law distribution is
$
\chi \sim \xi^d \sim \left| \Delta p\right|^{-d\nu},
$
such that
$\sigma=\frac{1}{d \nu}=\frac{3}{8}$, since $\chi \sim \left| \Delta p\right|^{-1/\sigma} $,
For $ \tau<2 $ the average cluster size,
\begin{equation}
S=\frac{1}{M}\sum_{s=1}^{\infty}{}^{'}n_{s}s^2,
\label{Sdef}
\end{equation}
scales as the cutoff, $ \chi $, such that
$\gamma=\frac{1}{\sigma}=\frac{8}{3}$, since $S \sim \left| \Delta p\right|^{-\gamma}$.
The prime in Eq.~\eqref{Sdef} indicates that the sum runs over all non-percolating clusters.
We expect the following scaling ansatz for the average cluster mass (see SI)
\begin{equation}
S = M \mathcal{H} \left( \Delta p M^{1/\gamma}\right).
\label{Scollapse}
\end{equation}

The quantity $S$ in Eq.\ \eqref{Sdef} is analogous to the susceptibility in thermal systems, where the diverging susceptibility is a signature of a \textit{second}-order phase transition. In the NEP model the divergence of $S$ comes along with a discontinuity of $P$---a signature of the \textit{first}-order-like phase transition. Therefore, the critical exponent $ \beta $ and the conductivity exponent $ \mu $ are both zero, such that the usual scaling law $ \gamma + 2\beta = d\nu $ remains valid.

As one can see from Eqs.\ (\ref{ns},\ref{nsP}), the actual value of the Fisher exponent is determined by the functional dependence of $\mathcal{P}\left(s\right)$. Since the only relevant mass scale at the transition point of the RP model is $M^{d_\mathrm{f}'/d}$,
\begin{equation}
\mathcal{P}\left(s\right)=\mathcal{F}\left(\frac{s}{M^{d_\mathrm{f}'/d}}\right) \sim \left(\frac{s}{M^{d_\mathrm{f}'/d}}\right)^{a}, 
\label{P}
\end{equation}
where $\mathcal{F}$ is a scaling function and (after combining Eqs.\ (\ref{ns},\ref{nsP},\ref{P}))
\begin{equation}
a=\left(2-\tau\right)\frac{d}{d_\mathrm{f}'}.
\label{a}
\end{equation}
Other properties of the NEP model are derived and summarized in SI

\textit{Numerical analysis}---To verify our theoretical analysis of the NEP model, we simulate it on a 2D triangular lattice (see details in SI), where a random occupation of bonds results in a continuous, second-order like, transition when the probability of a bond to be occupied is $ p=p_c=2\sin \left(\pi/18 \right) $ \cite{essam1978percolation}. For the NEP model, we first demonstrate the discontinuous transition behavior of $P$. As shown in Fig.\ \ref{NEP4}(b), the transition of $P$ from $0$ to $1$ as a function of $p$ becomes steeper with increasing system size, $M$.
In fact, as shown in inset of Fig.\ \ref{NEP4}(b), one can collapse the data from Fig.\ \ref{NEP4} using the scaling form \eqref{Pcollapse} with the calculated value of $\gamma=8/3$. Therefore, in the limit $M \rightarrow \infty$ at $ p = p_c$ the value of $P$ discontinuously jumps from zero to one.

\/*
\begin{figure}[b]
\centering
    \includegraphics[width=\columnwidth]{Pfig.eps} 
\caption{Demonstration of the discontinuous transition in the NEP model. Strength of the largest cluster, $P$, as a function of bond density, $p$ for different system sizes (see legend). Inset: collapse of the data from the main figure using Eq.\ \eqref{Pcollapse} and $\gamma=8/3$.}
\label{Pfig}
\end{figure}
*/

To demonstrate criticality in the NEP model, we perform finite-size analysis of the average, non-percolating cluster size---the analogue of the susceptibility in thermal systems, plotting $ S $ vs $ p $ for different values of the total mass of the system, $ M $. In Fig.\ \ref{NEP4}(c) one can see that the peak value of $ S $ depends on the system size. In fact, as shown in inset of Fig.\ \ref{NEP4}(c), one can collapse the data from Fig. \ref{NEP4}(c) using the scaling form \eqref{Scollapse} with the calculated value of $\gamma=8/3$. Therefore, in the limit $M \rightarrow \infty$, $ S $ diverges at the transition, as in a second-order phase transition.  Thus, our finite-size scaling analysis confirms the hybrid nature of the phase transition in the NEP model, with discontinuity in the order parameter, $ P $, and yet divergence of the susceptibility, $ S $, in the thermodynamic limit.

As is mentioned above, such a hybrid phase transition can exist only if the Fisher exponent is smaller than its an upper bound of $2$. As one can see in Fig.\ \ref{NEP4}(a) the cluster mass distribution is consistent with Eq.\ \eqref{ns} with
\begin{equation}
\tau=1.82 \pm 0.01<2,
\label{tau}
\end{equation}
in agreement with the considerations above. The above estimation of $\tau$ in Eq.\ \eqref{tau} was obtained using Eq.\ \eqref{n} (see details in SI). The obtained value of $\tau$ is in agreement with our experimental results (see Fig. \ref{Jose}(c)) 
 
The difference of $\tau$ from $2$ is due to a power-law scaling of the $\mathcal{P}\left( s \right) \sim \left(s/M^{d_\mathrm{f}'/d}\right)^a$, as indicated in Eq.\ \eqref{a}. To verify this, we calculated numerically the probability of an $s$-cluster in the RP model to be not an enclave for different systems sizes, $M$. As shown in Fig.\ \ref{NEP4}(d) the dependence is consistent with Eqs.\ (\ref{P},\ref{a},\ref{tau}). Moreover, all the $\mathcal{P}\left(s\right)$ data, for different $M$ values, collapses to a master curve with a power-law dependence on $s/M^{d_\mathrm{f}'/d}$, confirming Eq.\ \eqref{P} (see inset of Fig.\ \ref{NEP4}(d)). With this we complete the numerical validation of the analytical arguments.


We have shown that a simple extension of the RP model, with absorbed enclaves, exhibits simultaneously features of both discontinuous and continuous phase transitions. This model is in a distinct universality class from the RP model, with different critical exponents for all but the correlation length. Importantly, motor activated gels appear to constitute an experimental realization of this universality class.

\acknowledgments
This work was supported by FOM/NWO. The authors thank D.\ Stauffer,
Y.\ Kafri and R.\ Ziff for helpful discussions.

\bibliographystyle{apsrev}
\bibliography{SOCbib}

\begin{thebibliography}{36}
\expandafter\ifx\csname natexlab\endcsname\relax\def\natexlab#1{#1}\fi
\expandafter\ifx\csname bibnamefont\endcsname\relax
  \def\bibnamefont#1{#1}\fi
\expandafter\ifx\csname bibfnamefont\endcsname\relax
  \def\bibfnamefont#1{#1}\fi
\expandafter\ifx\csname citenamefont\endcsname\relax
  \def\citenamefont#1{#1}\fi
\expandafter\ifx\csname url\endcsname\relax
  \def\url#1{\texttt{#1}}\fi
\expandafter\ifx\csname urlprefix\endcsname\relax\def\urlprefix{URL }\fi
\providecommand{\bibinfo}[2]{#2}
\providecommand{\eprint}[2][]{\url{#2}}

\bibitem[{\citenamefont{Sahimi}(1994)}]{sahimi1994applications}
\bibinfo{author}{\bibfnamefont{M.}~\bibnamefont{Sahimi}},
  \emph{\bibinfo{title}{Applications of percolation theory}}
  (\bibinfo{publisher}{CRC Press}, \bibinfo{year}{1994}).

\bibitem[{\citenamefont{De~Gennes}(1979)}]{de1979scaling}
\bibinfo{author}{\bibfnamefont{P.-G.} \bibnamefont{De~Gennes}},
  \emph{\bibinfo{title}{Scaling concepts in polymer physics}}
  (\bibinfo{publisher}{Cornell university press}, \bibinfo{year}{1979}).

\bibitem[{\citenamefont{Broedersz and
  MacKintosh}(2014)}]{broedersz2014modeling}
\bibinfo{author}{\bibfnamefont{C.}~\bibnamefont{Broedersz}} \bibnamefont{and}
  \bibinfo{author}{\bibfnamefont{F.}~\bibnamefont{MacKintosh}},
  \bibinfo{journal}{Rev. Mod. Phys.} \textbf{\bibinfo{volume}{86}},
  \bibinfo{pages}{995} (\bibinfo{year}{2014}).

\bibitem[{\citenamefont{Nagler et~al.}(2011)\citenamefont{Nagler, Levina, and
  Timme}}]{nagler2011impact}
\bibinfo{author}{\bibfnamefont{J.}~\bibnamefont{Nagler}},
  \bibinfo{author}{\bibfnamefont{A.}~\bibnamefont{Levina}}, \bibnamefont{and}
  \bibinfo{author}{\bibfnamefont{M.}~\bibnamefont{Timme}},
  \bibinfo{journal}{Nature Physics} \textbf{\bibinfo{volume}{7}},
  \bibinfo{pages}{265} (\bibinfo{year}{2011}).

\bibitem[{\citenamefont{Alvarado et~al.}(2013)\citenamefont{Alvarado, Sheinman,
  Sharma, MacKintosh, and Koenderink}}]{alvarado2013myosin}
\bibinfo{author}{\bibfnamefont{J.}~\bibnamefont{Alvarado}},
  \bibinfo{author}{\bibfnamefont{M.}~\bibnamefont{Sheinman}},
  \bibinfo{author}{\bibfnamefont{A.}~\bibnamefont{Sharma}},
  \bibinfo{author}{\bibfnamefont{F.}~\bibnamefont{MacKintosh}},
  \bibnamefont{and}
  \bibinfo{author}{\bibfnamefont{G.}~\bibnamefont{Koenderink}},
  \bibinfo{journal}{Nature Physics} \textbf{\bibinfo{volume}{9}},
  \bibinfo{pages}{591} (\bibinfo{year}{2013}).

\bibitem[{\citenamefont{Stauffer and Aharony}(1994)}]{stauffer1994introduction}
\bibinfo{author}{\bibfnamefont{D.}~\bibnamefont{Stauffer}} \bibnamefont{and}
  \bibinfo{author}{\bibfnamefont{A.}~\bibnamefont{Aharony}},
  \emph{\bibinfo{title}{Introduction To Percolation Theory}}
  (\bibinfo{publisher}{Taylor \& Francis}, \bibinfo{year}{1994}).

\bibitem[{\citenamefont{Den~Nijs}(1979)}]{den1979relation}
\bibinfo{author}{\bibfnamefont{M.}~\bibnamefont{Den~Nijs}},
  \bibinfo{journal}{Journal of Physics A: Mathematical and General}
  \textbf{\bibinfo{volume}{12}}, \bibinfo{pages}{1857} (\bibinfo{year}{1979}).

\bibitem[{\citenamefont{Nienhuis}(1982)}]{nienhuis1982exact}
\bibinfo{author}{\bibfnamefont{B.}~\bibnamefont{Nienhuis}},
  \bibinfo{journal}{Phys. Rev. Lett.} \textbf{\bibinfo{volume}{49}},
  \bibinfo{pages}{1062} (\bibinfo{year}{1982}).

\bibitem[{\citenamefont{Clauset et~al.}(2009)\citenamefont{Clauset, Shalizi,
  and Newman}}]{clauset2009power}
\bibinfo{author}{\bibfnamefont{A.}~\bibnamefont{Clauset}},
  \bibinfo{author}{\bibfnamefont{C.}~\bibnamefont{Shalizi}}, \bibnamefont{and}
  \bibinfo{author}{\bibfnamefont{M.}~\bibnamefont{Newman}},
  \bibinfo{journal}{SIAM review} \textbf{\bibinfo{volume}{51}},
  \bibinfo{pages}{661} (\bibinfo{year}{2009}).

\bibitem[{\citenamefont{Friedman and
  Landsberg}(2009)}]{Friedman2009Construction}
\bibinfo{author}{\bibfnamefont{E.~J.} \bibnamefont{Friedman}} \bibnamefont{and}
  \bibinfo{author}{\bibfnamefont{A.~S.} \bibnamefont{Landsberg}},
  \bibinfo{journal}{Phys. Rev. Lett.} \textbf{\bibinfo{volume}{103}},
  \bibinfo{pages}{255701} (\bibinfo{year}{2009}).

\bibitem[{\citenamefont{da~Costa et~al.}(2010)\citenamefont{da~Costa,
  Dorogovtsev, Goltsev, and Mendes}}]{da2010explosive}
\bibinfo{author}{\bibfnamefont{R.}~\bibnamefont{da~Costa}},
  \bibinfo{author}{\bibfnamefont{S.}~\bibnamefont{Dorogovtsev}},
  \bibinfo{author}{\bibfnamefont{A.}~\bibnamefont{Goltsev}}, \bibnamefont{and}
  \bibinfo{author}{\bibfnamefont{J.}~\bibnamefont{Mendes}},
  \bibinfo{journal}{Phys. Rev. Lett.} \textbf{\bibinfo{volume}{105}},
  \bibinfo{pages}{255701} (\bibinfo{year}{2010}).

\bibitem[{\citenamefont{Hooyberghs and
  Van~Schaeybroeck}(2011)}]{hooyberghs2011criterion}
\bibinfo{author}{\bibfnamefont{H.}~\bibnamefont{Hooyberghs}} \bibnamefont{and}
  \bibinfo{author}{\bibfnamefont{B.}~\bibnamefont{Van~Schaeybroeck}},
  \bibinfo{journal}{Phys. Rev. E} \textbf{\bibinfo{volume}{83}},
  \bibinfo{pages}{032101} (\bibinfo{year}{2011}).

\bibitem[{\citenamefont{da~Costa
  et~al.}(2014{\natexlab{a}})\citenamefont{da~Costa, Dorogovtsev, Goltsev, and
  Mendes}}]{da2014Critical}
\bibinfo{author}{\bibfnamefont{R.~A.} \bibnamefont{da~Costa}},
  \bibinfo{author}{\bibfnamefont{S.~N.} \bibnamefont{Dorogovtsev}},
  \bibinfo{author}{\bibfnamefont{A.~V.} \bibnamefont{Goltsev}},
  \bibnamefont{and} \bibinfo{author}{\bibfnamefont{J.~F.~F.}
  \bibnamefont{Mendes}}, \bibinfo{journal}{Phys. Rev. E}
  \textbf{\bibinfo{volume}{89}}, \bibinfo{pages}{042148}
  (\bibinfo{year}{2014}{\natexlab{a}}).

\bibitem[{\citenamefont{Buldyrev et~al.}(2010)\citenamefont{Buldyrev, Parshani,
  Paul, Stanley, and Havlin}}]{buldyrev2010catastrophic}
\bibinfo{author}{\bibfnamefont{S.}~\bibnamefont{Buldyrev}},
  \bibinfo{author}{\bibfnamefont{R.}~\bibnamefont{Parshani}},
  \bibinfo{author}{\bibfnamefont{G.}~\bibnamefont{Paul}},
  \bibinfo{author}{\bibfnamefont{H.}~\bibnamefont{Stanley}}, \bibnamefont{and}
  \bibinfo{author}{\bibfnamefont{S.}~\bibnamefont{Havlin}},
  \bibinfo{journal}{Nature} \textbf{\bibinfo{volume}{464}},
  \bibinfo{pages}{1025} (\bibinfo{year}{2010}).

\bibitem[{\citenamefont{Gao et~al.}(2011)\citenamefont{Gao, Buldyrev, Stanley,
  and Havlin}}]{gao2011networks}
\bibinfo{author}{\bibfnamefont{J.}~\bibnamefont{Gao}},
  \bibinfo{author}{\bibfnamefont{S.}~\bibnamefont{Buldyrev}},
  \bibinfo{author}{\bibfnamefont{H.}~\bibnamefont{Stanley}}, \bibnamefont{and}
  \bibinfo{author}{\bibfnamefont{S.}~\bibnamefont{Havlin}},
  \bibinfo{journal}{Nature Physics} \textbf{\bibinfo{volume}{8}},
  \bibinfo{pages}{40} (\bibinfo{year}{2011}).

\bibitem[{\citenamefont{Boettcher et~al.}(2012)\citenamefont{Boettcher, Singh,
  and Ziff}}]{boettcher2012ordinary}
\bibinfo{author}{\bibfnamefont{S.}~\bibnamefont{Boettcher}},
  \bibinfo{author}{\bibfnamefont{V.}~\bibnamefont{Singh}}, \bibnamefont{and}
  \bibinfo{author}{\bibfnamefont{R.}~\bibnamefont{Ziff}},
  \bibinfo{journal}{Nature Communications} \textbf{\bibinfo{volume}{3}},
  \bibinfo{pages}{787} (\bibinfo{year}{2012}).

\bibitem[{\citenamefont{Achlioptas et~al.}(2009)\citenamefont{Achlioptas,
  D'Souza, and Spencer}}]{achlioptas2009explosive}
\bibinfo{author}{\bibfnamefont{D.}~\bibnamefont{Achlioptas}},
  \bibinfo{author}{\bibfnamefont{R.}~\bibnamefont{D'Souza}}, \bibnamefont{and}
  \bibinfo{author}{\bibfnamefont{J.}~\bibnamefont{Spencer}},
  \bibinfo{journal}{Science} \textbf{\bibinfo{volume}{323}},
  \bibinfo{pages}{1453} (\bibinfo{year}{2009}).

\bibitem[{\citenamefont{Moreira et~al.}(2010)\citenamefont{Moreira, Oliveira,
  Reis, Herrmann, and Andrade~Jr}}]{moreira2010hamiltonian}
\bibinfo{author}{\bibfnamefont{A.}~\bibnamefont{Moreira}},
  \bibinfo{author}{\bibfnamefont{E.}~\bibnamefont{Oliveira}},
  \bibinfo{author}{\bibfnamefont{S.}~\bibnamefont{Reis}},
  \bibinfo{author}{\bibfnamefont{H.}~\bibnamefont{Herrmann}}, \bibnamefont{and}
  \bibinfo{author}{\bibfnamefont{J.}~\bibnamefont{Andrade~Jr}},
  \bibinfo{journal}{Phys. Rev. E} \textbf{\bibinfo{volume}{81}},
  \bibinfo{pages}{040101} (\bibinfo{year}{2010}).

\bibitem[{\citenamefont{Cho et~al.}(2010)\citenamefont{Cho, Kahng, and
  Kim}}]{cho2010cluster}
\bibinfo{author}{\bibfnamefont{Y.~S.} \bibnamefont{Cho}},
  \bibinfo{author}{\bibfnamefont{B.}~\bibnamefont{Kahng}}, \bibnamefont{and}
  \bibinfo{author}{\bibfnamefont{D.}~\bibnamefont{Kim}},
  \bibinfo{journal}{Phys. Rev. E} \textbf{\bibinfo{volume}{81}},
  \bibinfo{pages}{030103} (\bibinfo{year}{2010}).

\bibitem[{\citenamefont{D'Souza and Mitzenmacher}(2010)}]{d2010local}
\bibinfo{author}{\bibfnamefont{R.~M.} \bibnamefont{D'Souza}} \bibnamefont{and}
  \bibinfo{author}{\bibfnamefont{M.}~\bibnamefont{Mitzenmacher}},
  \bibinfo{journal}{Phys. Rev. Lett.} \textbf{\bibinfo{volume}{104}},
  \bibinfo{pages}{195702} (\bibinfo{year}{2010}).

\bibitem[{\citenamefont{Radicchi and Fortunato}(2010)}]{radicchi2010explosive}
\bibinfo{author}{\bibfnamefont{F.}~\bibnamefont{Radicchi}} \bibnamefont{and}
  \bibinfo{author}{\bibfnamefont{S.}~\bibnamefont{Fortunato}},
  \bibinfo{journal}{Phys. Rev. E} \textbf{\bibinfo{volume}{81}},
  \bibinfo{pages}{036110} (\bibinfo{year}{2010}).

\bibitem[{\citenamefont{Ara{\'u}jo and Herrmann}(2010)}]{araujo2010explosive}
\bibinfo{author}{\bibfnamefont{N.}~\bibnamefont{Ara{\'u}jo}} \bibnamefont{and}
  \bibinfo{author}{\bibfnamefont{H.}~\bibnamefont{Herrmann}},
  \bibinfo{journal}{Phys. Rev. Lett.} \textbf{\bibinfo{volume}{105}},
  \bibinfo{pages}{035701} (\bibinfo{year}{2010}).

\bibitem[{\citenamefont{Riordan and Warnke}(2011)}]{riordan2011explosive}
\bibinfo{author}{\bibfnamefont{O.}~\bibnamefont{Riordan}} \bibnamefont{and}
  \bibinfo{author}{\bibfnamefont{L.}~\bibnamefont{Warnke}},
  \bibinfo{journal}{Science} \textbf{\bibinfo{volume}{333}},
  \bibinfo{pages}{322} (\bibinfo{year}{2011}).

\bibitem[{\citenamefont{Nagler et~al.}(2012)\citenamefont{Nagler, Tiessen, and
  Gutch}}]{nagler2012continuous}
\bibinfo{author}{\bibfnamefont{J.}~\bibnamefont{Nagler}},
  \bibinfo{author}{\bibfnamefont{T.}~\bibnamefont{Tiessen}}, \bibnamefont{and}
  \bibinfo{author}{\bibfnamefont{H.~W.} \bibnamefont{Gutch}},
  \bibinfo{journal}{Phys. Rev. X} \textbf{\bibinfo{volume}{2}},
  \bibinfo{pages}{031009} (\bibinfo{year}{2012}).

\bibitem[{\citenamefont{da~Costa
  et~al.}(2014{\natexlab{b}})\citenamefont{da~Costa, Dorogovtsev, Goltsev, and
  Mendes}}]{da2014Solution}
\bibinfo{author}{\bibfnamefont{R.~A.} \bibnamefont{da~Costa}},
  \bibinfo{author}{\bibfnamefont{S.~N.} \bibnamefont{Dorogovtsev}},
  \bibinfo{author}{\bibfnamefont{A.~V.} \bibnamefont{Goltsev}},
  \bibnamefont{and} \bibinfo{author}{\bibfnamefont{J.~F.~F.}
  \bibnamefont{Mendes}}, \bibinfo{journal}{Phys. Rev. E}
  \textbf{\bibinfo{volume}{90}}, \bibinfo{pages}{022145}
  (\bibinfo{year}{2014}{\natexlab{b}}).

\bibitem[{\citenamefont{Ara{\'u}jo et~al.}(2011)\citenamefont{Ara{\'u}jo,
  Andrade~Jr, Ziff, and Herrmann}}]{araujo2011tricritical}
\bibinfo{author}{\bibfnamefont{N.~A.} \bibnamefont{Ara{\'u}jo}},
  \bibinfo{author}{\bibfnamefont{J.~S.} \bibnamefont{Andrade~Jr}},
  \bibinfo{author}{\bibfnamefont{R.~M.} \bibnamefont{Ziff}}, \bibnamefont{and}
  \bibinfo{author}{\bibfnamefont{H.~J.} \bibnamefont{Herrmann}},
  \bibinfo{journal}{Phys. Rev. Lett.} \textbf{\bibinfo{volume}{106}},
  \bibinfo{pages}{095703} (\bibinfo{year}{2011}).

\bibitem[{\citenamefont{Cho et~al.}(2013)\citenamefont{Cho, Hwang, Herrmann,
  and Kahng}}]{cho2013avoiding}
\bibinfo{author}{\bibfnamefont{Y.}~\bibnamefont{Cho}},
  \bibinfo{author}{\bibfnamefont{S.}~\bibnamefont{Hwang}},
  \bibinfo{author}{\bibfnamefont{H.}~\bibnamefont{Herrmann}}, \bibnamefont{and}
  \bibinfo{author}{\bibfnamefont{B.}~\bibnamefont{Kahng}},
  \bibinfo{journal}{Science} \textbf{\bibinfo{volume}{339}},
  \bibinfo{pages}{1185} (\bibinfo{year}{2013}).

\bibitem[{\citenamefont{Schr{\"o}der et~al.}(2013)\citenamefont{Schr{\"o}der,
  Rahbari, and Nagler}}]{schroder2013crackling}
\bibinfo{author}{\bibfnamefont{M.}~\bibnamefont{Schr{\"o}der}},
  \bibinfo{author}{\bibfnamefont{S.~E.} \bibnamefont{Rahbari}},
  \bibnamefont{and} \bibinfo{author}{\bibfnamefont{J.}~\bibnamefont{Nagler}},
  \bibinfo{journal}{Nature communications} \textbf{\bibinfo{volume}{4}}
  (\bibinfo{year}{2013}).

\bibitem[{\citenamefont{Chalupa et~al.}(1979)\citenamefont{Chalupa, Leath, and
  Reich}}]{chalupa1979bootstrap}
\bibinfo{author}{\bibfnamefont{J.}~\bibnamefont{Chalupa}},
  \bibinfo{author}{\bibfnamefont{P.~L.} \bibnamefont{Leath}}, \bibnamefont{and}
  \bibinfo{author}{\bibfnamefont{G.~R.} \bibnamefont{Reich}},
  \bibinfo{journal}{Journal of Physics C: Solid State Physics}
  \textbf{\bibinfo{volume}{12}}, \bibinfo{pages}{L31} (\bibinfo{year}{1979}).

\bibitem[{\citenamefont{Schwarz et~al.}(2006)\citenamefont{Schwarz, Liu, and
  Chayes}}]{schwarz2006onset}
\bibinfo{author}{\bibfnamefont{J.}~\bibnamefont{Schwarz}},
  \bibinfo{author}{\bibfnamefont{A.~J.} \bibnamefont{Liu}}, \bibnamefont{and}
  \bibinfo{author}{\bibfnamefont{L.}~\bibnamefont{Chayes}},
  \bibinfo{journal}{EPL (Europhysics Letters)} \textbf{\bibinfo{volume}{73}},
  \bibinfo{pages}{560} (\bibinfo{year}{2006}).

\bibitem[{\citenamefont{Chae et~al.}(2014)\citenamefont{Chae, Yook, and
  Kim}}]{Chae2014Complete}
\bibinfo{author}{\bibfnamefont{H.}~\bibnamefont{Chae}},
  \bibinfo{author}{\bibfnamefont{S.-H.} \bibnamefont{Yook}}, \bibnamefont{and}
  \bibinfo{author}{\bibfnamefont{Y.}~\bibnamefont{Kim}},
  \bibinfo{journal}{Phys. Rev. E} \textbf{\bibinfo{volume}{89}},
  \bibinfo{pages}{052134} (\bibinfo{year}{2014}).

\bibitem[{\citenamefont{Thouless}(1969)}]{thouless1969long}
\bibinfo{author}{\bibfnamefont{D.~J.} \bibnamefont{Thouless}},
  \bibinfo{journal}{Phys. Rev.} \textbf{\bibinfo{volume}{187}},
  \bibinfo{pages}{732} (\bibinfo{year}{1969}).

\bibitem[{\citenamefont{Kafri et~al.}(2000)\citenamefont{Kafri, Mukamel, and
  Peliti}}]{kafri2000why}
\bibinfo{author}{\bibfnamefont{Y.}~\bibnamefont{Kafri}},
  \bibinfo{author}{\bibfnamefont{D.}~\bibnamefont{Mukamel}}, \bibnamefont{and}
  \bibinfo{author}{\bibfnamefont{L.}~\bibnamefont{Peliti}},
  \bibinfo{journal}{Phys. Rev. Lett.} \textbf{\bibinfo{volume}{85}},
  \bibinfo{pages}{4988} (\bibinfo{year}{2000}).

\bibitem[{\citenamefont{Bar and Mukamel}(2014)}]{bar2014mixed}
\bibinfo{author}{\bibfnamefont{A.}~\bibnamefont{Bar}} \bibnamefont{and}
  \bibinfo{author}{\bibfnamefont{D.}~\bibnamefont{Mukamel}},
  \bibinfo{journal}{Phys. Rev. Lett.} \textbf{\bibinfo{volume}{112}},
  \bibinfo{pages}{015701} (\bibinfo{year}{2014}).

\bibitem[{\citenamefont{Van~Hecke}(2010)}]{van2010jamming}
\bibinfo{author}{\bibfnamefont{M.}~\bibnamefont{Van~Hecke}},
  \bibinfo{journal}{Journal of Physics: Condensed Matter}
  \textbf{\bibinfo{volume}{22}}, \bibinfo{pages}{033101}
  (\bibinfo{year}{2010}).

\bibitem[{\citenamefont{Essam et~al.}(1978)\citenamefont{Essam, Gaunt, and
  Guttmann}}]{essam1978percolation}
\bibinfo{author}{\bibfnamefont{I.}~\bibnamefont{Essam}},
  \bibinfo{author}{\bibfnamefont{D.}~\bibnamefont{Gaunt}}, \bibnamefont{and}
  \bibinfo{author}{\bibfnamefont{A.}~\bibnamefont{Guttmann}},
  \bibinfo{journal}{J. Physics A: Mathematical and General}
  \textbf{\bibinfo{volume}{11}}, \bibinfo{pages}{1983} (\bibinfo{year}{1978}).

\end{thebibliography}


\clearpage
\begin{center}
\Huge
Supplementary Material
\end{center}

\appendix

\setcounter{figure}{0}
\setcounter{equation}{0}
\renewcommand{\theequation}{S\arabic{equation}}

\section*{Experimental results}
Experimentally, we study the motor-driven collapse of model cytoskeletal system, composed of actin filaments, fascin cross-links and myosin motors, using the approach, developed in Ref.~\cite{alvarado2013myosin}.
The components of the network were injected into a quasi-2D chamber of approximated dimensions $3mm \times 2mm \times 80 \mu m$. The thickness of the system ($80 \mu m$) is much smaller than the size of its other two dimensions and is comparable to the length of the longest actin filaments ($20 \mu m$), making the system, effectively, two-dimensional. Cluster sizes were measured by recording images of contracting actomyosin networks and tracking cluster expansion in reverse time using a customized image analysis algorithm. Cluster size distributions were determined for three sample regimes (controlled by crosslink density, provided that the motor density is kept constant): many small clusters, clusters with scale-free distributed sizes, and few large clusters.

In Fig. \ref{Jose} and the movie one can see experimental results of a fascin-crosslinked actin network, collapsed by myosin motors during $104$ min. The concentrations of crosslinks, actin and molecular motors are given by $0.24$, $12$ and $0.12$ $\mu M$, respectively. 

\/*
We find that the second regime surprisingly spans a wide range of experimental parameters and is manifested in a scale-free cluster size distribution:
\begin{equation}
n_s \sim s^{-\tau},
\label{ns1}
\end{equation}
as has been observed experimentally in Ref.~\cite{alvarado2013myosin}.
Here $n_s$ is the number of clusters with initial area (for a compact cluster its area is proportional to its mass) $s$ and $\tau$ is the Fisher power-law exponent.
This observation suggests a connectivity percolation phenomenon, where the clusters are also distributed in a scale-free manner at the percolation point. However, the details of the experimental results cannot be explained by a \textit{random} percolation model.

Firstly, as indicated in Ref.~\cite{alvarado2013myosin}, the robustness of the scale-free collapsed structure does not exist in a RP model, where the criticality is apparent only in a very narrow vicinity near the percolation point.
Secondly, what was not pointed out in Ref.~\cite{alvarado2013myosin}, the properties of the scale-free structure in the experiment and in a RP model are qualitatively different. In the RP model, at the percolation transition, clusters possess many scale-free enclaves---clusters fully surrounded by another cluster (see Fig. \ref{Enclaves} for illustration of enclaves). The presence of enclaves makes the clusters non-compact. In fact, the clusters are so porous, that their mass scales sublinearly with their volume. Such structures are random fractals, with the fractal dimension $d_\mathrm{f}'=91/48<2$. Since the density of the largest cluster vanishes in the thermodynamic limit, the percolation transition  is continuous \cite{stauffer1994introduction}. An example of a network configuration at the critical point of the RP model with such fractal clusters can be seen in Fig. \ref{InitialAreas}(a).
However, the properties of the scale-free structure in the experiment are different from those at the percolation transition point of the RP model. In the experiment the motors collapse the network to multiple small clusters (see Fig. \ref{Jose}(a,b) and movie in SI). The initial configuration of the collapsed clusters possesses no enclaves, as shown in Fig. \ref{Jose}(c). The absence of enclaves in the experiment suggests non-fractal, compact, euclidean clusters.
Moreover, the Fisher exponent, $\tau$, of the clusters' initial area distribution in Eq.~\eqref{ns1} in the experiment is smaller than $2$ (see Fig. \ref{Jose}(d)). This is inconsistent with a RP model. There the order parameter is continuous at the percolation transition, implying $\tau'=187/91>2$ \cite{stauffer1994introduction}. In fact, the Fisher exponent and fractal dimension are related via a hyperscaling relation \cite{stauffer1994introduction}. Therefore, it is expected that violation of $d_\mathrm{f}<2$ (as shown for the experimental system in Fig. \ref{Jose}(c)) also implies violation of $\tau>2$ (as shown for the experimental system in Fig. \ref{Jose}(d)).
*/

\section*{Absorption of enclave during the collapse}
\begin{figure}[!]
\renewcommand{\thefigure}{S\arabic{figure}}
\centering
\includegraphics[width= \columnwidth]{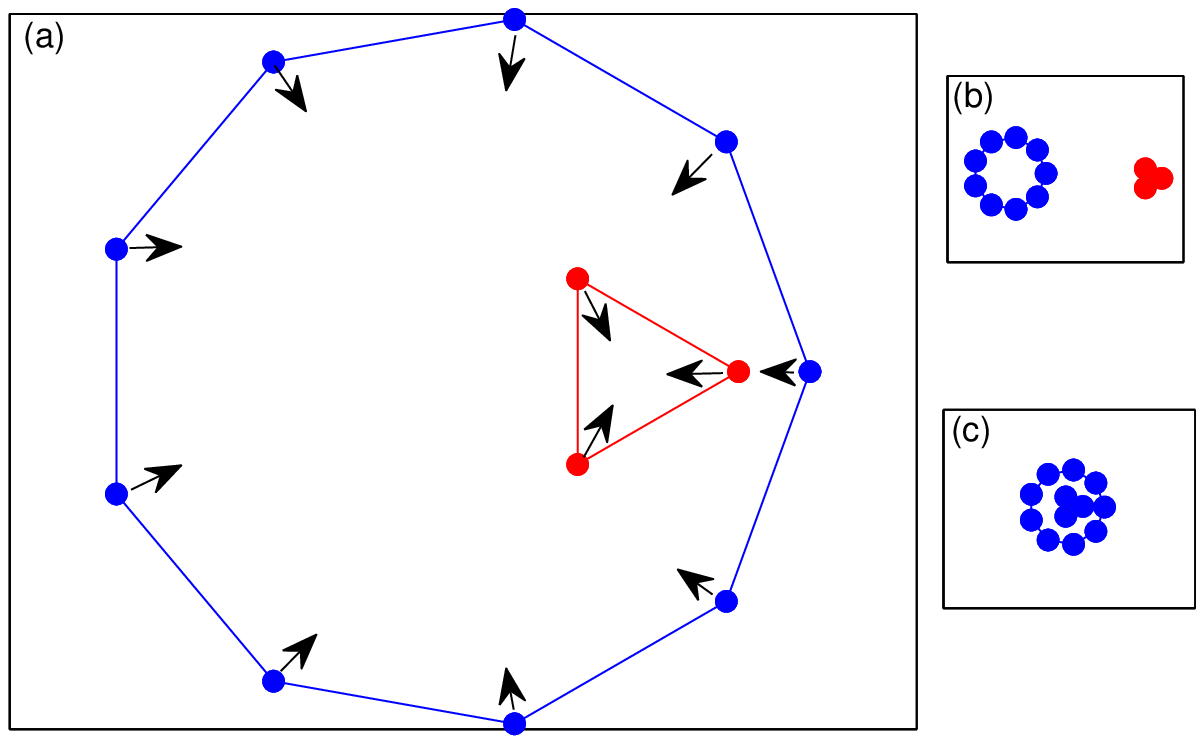}
\caption{An illustration of the influence of steric interactions on the collapsed state properties. The initial state presented in (a) without steric interactions collapses to two clusters with masses $ 3 $ and $ 9 $, as show in (b). Due to steric interactions the enclave is trapped in its surroundings, leading to a collapsed state with one cluster of mass $ 12 $, as shown in (c).}
\label{EnclaveInclusion}
\end{figure}

The absence of enclaves is expected from the following reasons. During the collapse process of active networks, due to excluded volume interactions, the enclaves of a cluster collapse to the same point as their surrounding cluster. In the experiment the initial configurations of the clusters were reconstructed, starting from the collapsed state. The collapsed state of a large cluster contains all its enclaves (see Fig. \ref{EnclaveInclusion} for illustration). This makes the reconstructed initial configuration of all clusters enclave-free (see Fig. \ref{Jose}(b)). This suggests that a RP model, possessing many enclaves, cannot have the same features as the collapsed state. In contrast, as we show in the main text, the NEP model where all the enclaves are absorbed, similarly to experimental system, accounts for its statistical properties. 

\section*{Numerical realization and illustration of the NEP model}
\begin{figure}[!] 
\renewcommand{\thefigure}{S\arabic{figure}}
\centering
  \begin{tabular}{@{}cccc@{}}
      \includegraphics[width=\columnwidth]{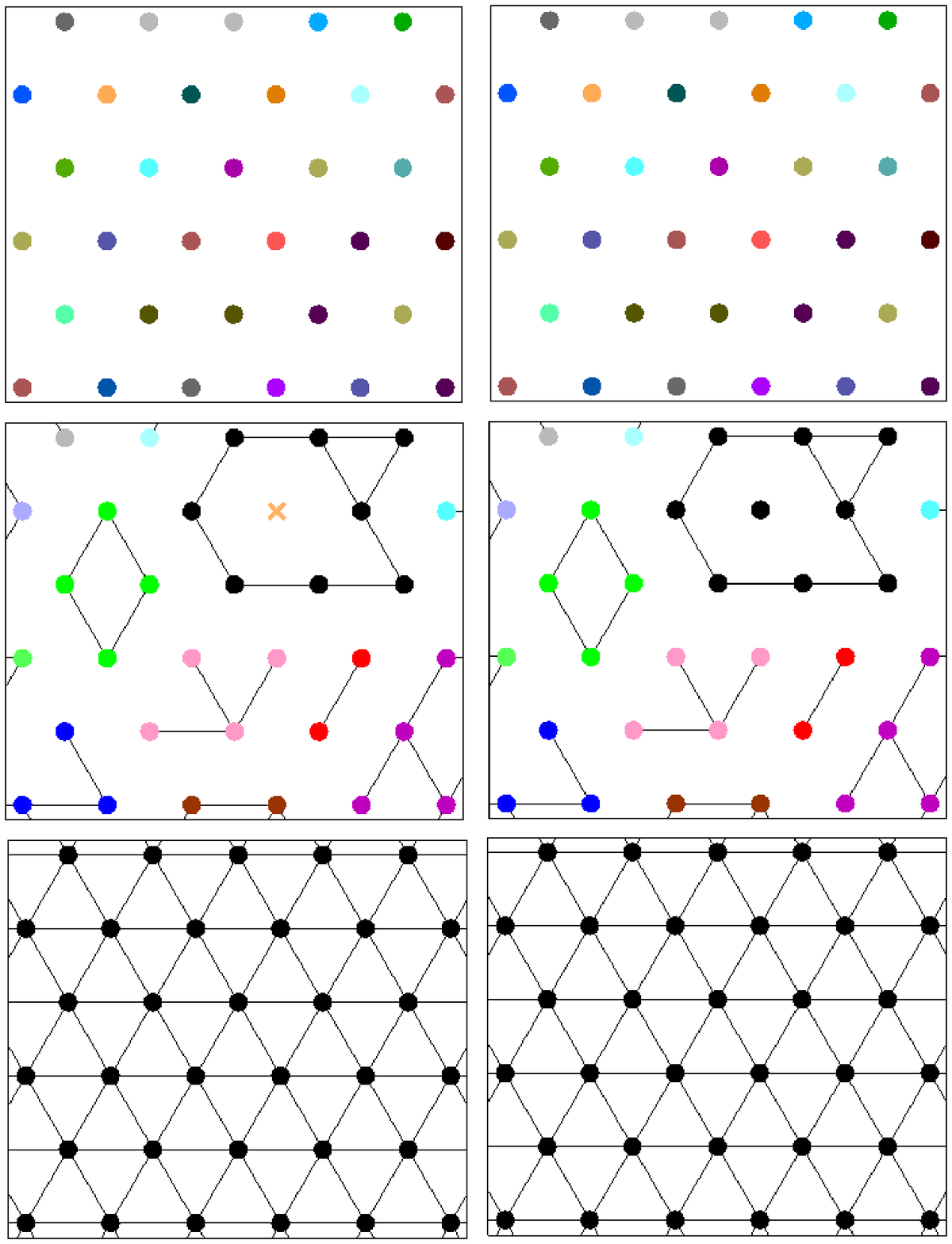} \\
    \includegraphics[width=.24\textwidth]{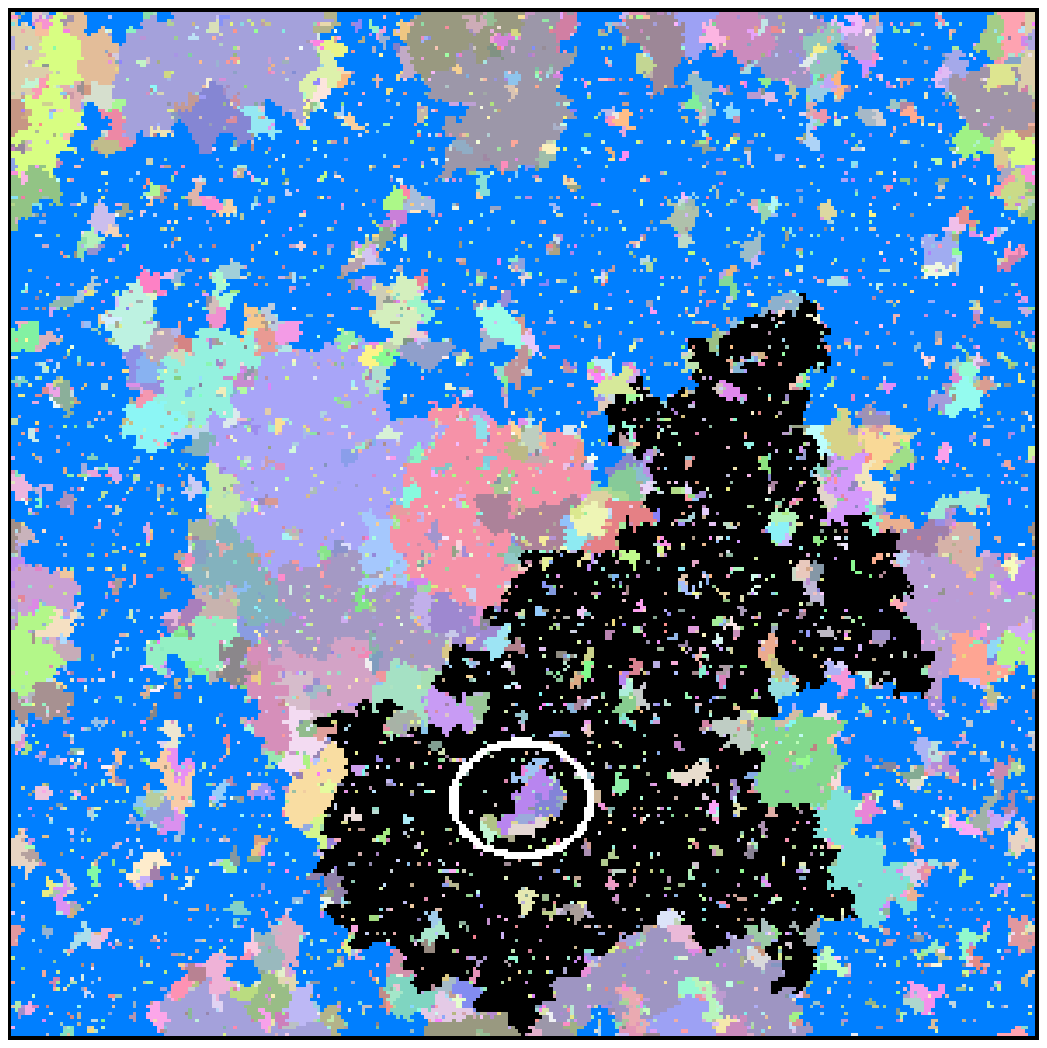} 
    \includegraphics[width=.24\textwidth]{Areas4-0.eps} 
  \end{tabular}
\caption{Cluster structure of the RP model (left panels) and the NEP model (right panels) on a triangular lattice for different values of $p$: $p=0$ (first two panels from top), $0<p<1$ (second two panels), $p=1$ (third two panels). Cross indicates an enclavic cluster which is absorbed to its surrounding in the NEP model. The bottom two panels depict cluster structure at the critical point, $p=p_c$. White ellipse in the left panel indicates an enclave, absorbed in the right panel. In every panel each cluster is indicated by a different color on each plot.}
\label{Picture}
\end{figure}

To get a configuration of clusters for the NEP model at a certain value of $p$ we, first, construct random percolation model with the same value of $p$. After this for each cluster we find its longest boundary and identify all the clusters within this boundary (enclaves) within the cluster. A boundary is defined as a collection of nodes such that they all are neighbors of the cluster and form a closed chain with a link of lattice constant. We note that the absorption of enclaves does not change the topology of the network. Therefore, the absorption can be performed at each step of the network dilution protocol. It follows that the resulting configuration of the network (with absorbed enclaves) is independent of the update protocol. An illustration of this process can be seen in Fig.\ \ref{Picture}.

\section*{Why $\tau<2$ is required for a phase transition with both criticality and discontinuity}
\label{App0}
In the main text we claim that $\tau<2$ is required for the mixed type of phase transition, with both criticality and discontinuity, found in the NEP model. The criticality follows from the existence of power-law distribution for any value of $\tau$. However, the discontinuity in the largest cluster's strength, $P$, demands that at the critical point the largest cluster's mass scales as the total mass. This condition is satisfied only for a heavy tail distribution, with $\tau<2$ at the critical point. Only in this case the prefactor of the cluster mass distribution is sublinear with $M$,
\begin{equation}
n_s \sim M^{\tau-1}s^{-\tau}.
\end{equation}
The largest cluster, $s_1$, can be calculated as
\begin{equation}
n_{s_1} s_1 \sim 1,
\end{equation}
resulting in $s_1 \sim M$, in contrast to case of $\tau>2$ where $s_1 \sim M^{d_{\rm f}/d} \sim M^{\frac{1}{\tau-1}}$ is sublinear with $M$, giving rise to continuous transition. See more in Refs.~\cite{Friedman2009Construction,da2010explosive,hooyberghs2011criterion,da2014Critical}.

\section*{Scaling ansatz for $ P $ and $ S $ }
\label{AppA}
In this section we rationalize Eqs. (\ref{Pcollapse}) and (\ref{Scollapse}). 
In contrast to the order in the main text, for clarity, we consider first the ansatz for $S$. The ansatz for $ S $ is based on the definition of $ \gamma $
\begin{equation}
S \sim \left| \Delta p\right|^{-\gamma}
\end{equation}
for $ M \rightarrow \infty $ and euclidean structure of the clusters,
\begin{equation}
S \sim M,
\end{equation}
at the transition point. This results in Eq. (\ref{Scollapse}).

The derivation of Eq. (\ref{Pcollapse}) is more subtle.
First, consider the case $ p<p_c $. In this case the clusters' mass is distributed as power-law with a cutoff $\chi$. The cutoff scales as $ S $, which scales as the largest cluster, $PM $, denoted here as $s_1$. 
Thus, one gets that
\begin{equation}
s_1 \sim S = M \mathcal{H} \left( \Delta p M^{1/\gamma}\right). 
\end{equation}Therefore, below the transition point this results in Eq. (\ref{Pcollapse}). Now we consider the case $p>p_c$.

For $ p>p_c $ the largest cluster's mass scales linearly with $ M $. The non-largest clusters's masses are distributed
as $ s^{-\tau} $ with a cutoff $ \chi$. Since, at $p=p_c$ the value of $ \chi $ scales linearly with $ M $ and the distribution is $n_s \sim M^{\tau-1}s^{-\tau}$, above $p_c$ the distribution of the nonspanning clusters' masses has to be of the form
\begin{equation}
n_s \sim \chi^{\tau-1} \left(\frac{M}{\chi}\right)^\alpha s^{-\tau}
\end{equation}
with $\alpha \leq 1$. The largest non-spanning cluster, $\chi$, in this case can be found using $n_{\chi}\chi \sim 1$ and is given by $\chi \sim M $ for any $\alpha>0$ and $\chi \sim M^0 $ for $\alpha=0$. Since above $p_c$ the cutoff as well as the correlation length is finite in the thermodynamic limit one gets that $\alpha$ has to be zero.

Therefore the total mass of the non-spanning clusters is given by (again, note that $ \tau<2 $)
\begin{equation}
\int_1^{\chi} n_s s ds=\chi^{\tau-1}\int_1^{\chi} s^{-\tau} s ds \sim \chi \sim S.
\end{equation}

Thus, since the spanning cluster possesses the total mass $ M $ minus the total mass of the non-spanning clusters, one obtains
\begin{equation}
P \sim 1-S/M
\end{equation}
leading to Eq. (\ref{Pcollapse}) also for $ p>p_c $. 

\section*{Other properties of the NEP model}
\label{OtherProperties}
\subsection*{Correlation function}
Absorption of enclaves also leads to constant correlation function within the clusters, $r \ll  \xi $ due to compactness of the clusters. Thus, the correlation function can be approximated as
\begin{equation}
G\left(r\right)\simeq\frac{\intop_{r^d}^{\xi^d}s n_{s}ds}{\intop_{1}^{\xi^d}s n_{s}ds}=\frac{\intop_{r^{d}}^{\xi^d}s^{1-\tau}ds}{\intop_{1}^{\xi^d}s^{1-\tau}ds}\simeq1-\left(\frac{r}{\xi}\right)^{\left(2-\tau\right)d}.
\end{equation}
Therefore, the anomalous dimension is given by $\eta=2-d=0$.
\subsection*{Relations between critical exponents}
Some relations between critical exponents have to be modified for the NEP model, with $ \tau<2 $, relative to the RP model with $\tau>2$:
\begin{equation}
\begin{array}{c|c}
\tau\geq2 & \tau\leq2\\
\hline \gamma=\frac{3-\tau}{\sigma} & \gamma=\frac{1}{\sigma}\\
\nu=\frac{\tau-1}{\sigma d} & \nu=\frac{1}{\sigma d}
\end{array} 
\end{equation}

\section*{Estimation of $\tau$ and summary of the NEP model's properties}
\label{AppB}
We estimate the value of $ \tau $ and the error bar of the estimation using Eq.~\eqref{ns}. We calculate numerically the number of clusters at the percolation point of the NEP model using two methods: directly (first one) and using the numerically calculated survival probability, $ \mathcal{P}(s) $, (second one): 
\begin{equation}
n = \sum_{s} n_s' \mathcal{P}(s) \sim M \sum_{s} s^{-\tau'} \mathcal{P}(s),
\label{nFromP}
\end{equation}
where $ n_s' $ is the number of $ s $-clusters in the RP model.

\begin{figure}[!]
\renewcommand{\thefigure}{S\arabic{figure}}
\centering
    \includegraphics[width=\columnwidth]{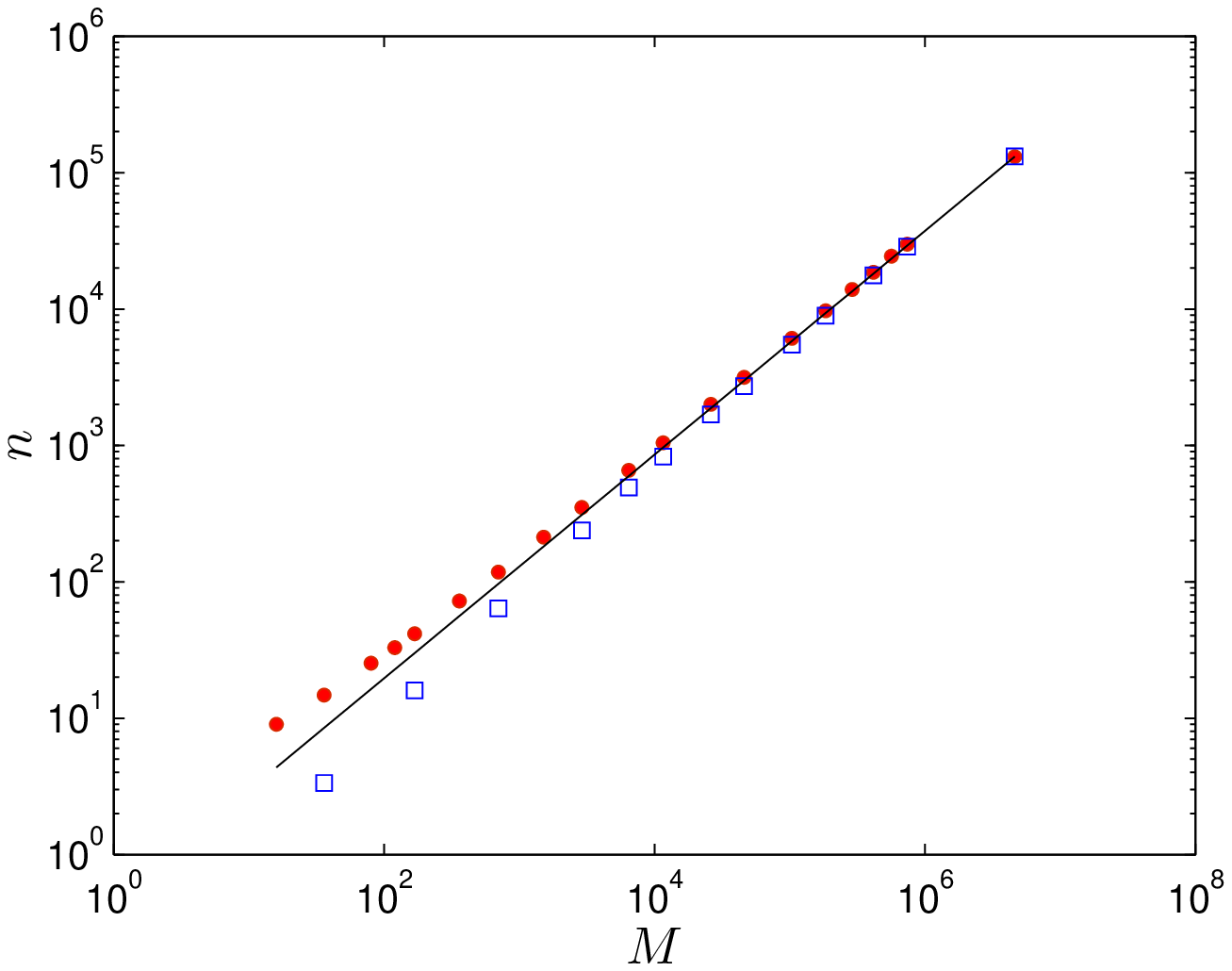} 
\caption{Number of clusters at the percolation transition of the NEP model as a function of the system size using two calculations: the direct one (squares) and using Eq.~\ref{nFromP} with a prefactor fitted to agree with the direct calculation for the largest $ M $ value (circles).}
\label{nVsM}
\end{figure}

In the large system limit the local slope of $ \log n $ vs. $ \log M $ equals to $ \tau-1 $ (see Eq.~\eqref{ns}). The local slopes of the two calculations converge to the same value in this limit, as shown in Fig. \ref{nVsM}. For finite $ M $ values the local slopes differ from the limiting value. The slope of the first calculation is larger and the second calculation is smaller than the limiting slope. This sandwich allows us to estimate the value of $ \tau $ as the average value of the local slopes for the largest system size. The confidence interval we estimate to be from the slope of the second calculation to the slope of the first one. This results in Eq.~\eqref{tau}. 

We summarize the results obtained from model in the following table.
\begin{equation}
\begin{array}{c|c|c}
 & \textrm{NEP model} & \textrm{RP}\\
\hline \tau & 1.82 \pm 0.01 & 187/91 \simeq 2.055\\
\nu & 4/3 & 4/3   \\
\gamma & 8/3  & 43/18 \simeq 2.389\\
d_\mathrm{f} & 2 & 91/48 \simeq 1.896 \\
\sigma & 3/8 = 0.375 & 36/91 \simeq 0.396\\
\beta & 0 & 5/36 \simeq 0.139\\
\eta & 0 & 5/24 \simeq 0.208\\
\Delta P & 1 & 0
\end{array}
\end{equation} 

\end{document}